\documentstyle{ioplppt}
\begin{document}
\jl{1}
\title{Statistical mechanics of the multi-constraint 
continuous knapsack problem}
\author{Jun-ichi Inoue}
\address{
Department of Physics,
Tokyo Institute of Technology,
Oh-okayama, Meguro-ku, Tokyo 152, Japan
}
\begin{abstract}
We apply the replica analysis established 
by  Gardner to the multi-constraint continuous knapsack problem,
which is one of the linear programming problems  and a most fundamental 
problem in the field of operations research (OR).
For a large problem size, we analyse 
the space of solution and its volume, and estimate the optimal 
number of items to go into the knapsack as a function of the number of 
constraints.
We study the stability of the replica symmetric (RS) solution
and find that the RS calculation cannot estimate the optimal 
number of items in knapsack correctly 
if many constraints are  required.
In order to obtain a consistent solution in the RS region,
we try the zero entropy approximation for this continuous 
solution space and get a stable solution within the RS ansatz.
On the other hand, in replica symmetry breaking (RSB) region, 
the one step RSB solution is found by Parisi's scheme. 
It turns out that this problem is closely related to the 
problem of optimal storage capacity 
and of generalization by maximum stability 
rule of a spherical perceptron.
\end{abstract}
\maketitle
\section{Introduction}
Recently, Korutcheva, Opper and L$\acute{\rm o}$pez \cite{Kou} 
pointed out 
that statistical mechanical analysis based on the replica 
calculation \cite{SK} can be used to investigate 
some specific statistical properties of optimal solutions for 
 an optimization problem, 
the so-called knapsack problem. 
The knapsack problem they studied is one of the integer 
programming problems in which the variables are all integer.
The problem is stated as follows.
Let us suppose 
that a man climbs a mountain. 
He puts $N$ items $s_{1},s_{2},{\cdots}
,s_{N}$ in his knapsack.
Each item  has 
its own weight 
$a_{1},a_{2},{\cdots},a_{N}$ and 
``utility'' $c_{1},c_{2},{\cdots},c_{N}$ 
which means, for example, its own ``price'' 
or ``necessity in climbing a mountain''.
He cannot bring all the $N$ items in  his  knapsack 
and he has  to  leave some of them.
He must decide 
which combination of items is best for him.
We can write down the above situation as 
\begin{eqnarray}
{\max}\left\{\sum_{j=1}^{N}c_{j}s_{j} \Bigm| \sum_{j=1}^{N}a_{j}s_{j}
{\leq}b,s_{j}{\in}\{0,1\},j=1,{\cdots}N \right\}
\end{eqnarray}
where the decision variable $s_{j}=1$ is defined such  
that the $j$th item goes into the knapsack, and $s_{j}=0$ otherwise. The constant $b$ represents the weight limit.
If we have to take account not only of the strength  of the 
man but also of the capacity  of the  knapsack itself ({\rm i.e.} 
the volume of the knapsack is limited), 
this new constraint must be considered and 
the problem becomes more complex.
Thus it is meaningful for us to generalize  our problem to a 
new form  with $K$ constraints as follows.\\
\begin{equation}
{\max}\left\{\sum_{j=1}^{N}c_{j}s_{j}\Bigm|\sum_{j=1}^{N}a_{kj}s_{j}
{\leq}b_{k},k=1,{\cdots},K,s_{j}{\in}\{0,1\},j=1,{\cdots},N \right \}
\end{equation}
Following Korutcheva \etal \cite{Kou},
we set the utilities
$c_{j}$ all $1/2$,
and we set the limits
$b_{k}$ all representing 
capacities 
of the man, 
of his knapsack, 
etc, to  $b$.

In order to treat the multi-constraint ($K{\sim}N$) knapsack 
problem in a more familiar 
way for physicists, 
we convert the decision variables $s_{j}=(1,0)$
into spin variables 
$\hat{s_{j}}=(1,-1)$, by transformation 
$\hat{s_{j}}=2(s_{j}-1/2)$.
The load $a_{kj}$ is also transformed as $a_{kj}=1/2+{\xi}_{kj}$, where the quenched random variable ${\xi}_{kj}$ has 
zero mean and variance ${\sigma}^{2}$
and is assumed to obey the distribution
\begin{eqnarray}
P({\xi}_{ki})=\frac{1}{\sqrt{2\pi}\sigma}{\exp}\left(-\frac{({\xi}_{ki})^{2}}
{2{\sigma}^{2}}\right)
\end{eqnarray}
As loads $a_{ki}$, 
which means ``weights" 
for example, should 
be positive, we must 
choose the above variance 
${\sigma}^{2}$ 
small enough and for this small variance, the fraction of 
items for which $a_{ki}<0$ for all $k$ vanishes exponentially with 
increasing $K$. 
In this paper, we use 
${\sigma}=1/12$ following Korutcheva {\it et.al} \cite{Kou}. 
Using these transformations, 
our problem is rewritten 
as follows. \\
\begin{equation}
{\max}\left\{U=\frac{N}{4}+\frac{1}{2}\sum_{j=1}^{N}\hat{s_{j}}\Bigm|
Y(\{{\bf s}\},\{a_{ki}\},b){\leq}0,
\mbox{}k=1,{\cdots},K,\hat{s_{j}}{\in}
\{1,-1\},j=1,{\cdots},N \right\}
\end{equation}
where
\begin{equation}
Y(\{{\bf s}\},\{a_{ki}\},b)
=\frac{1}{2}\sum_{j=1}^{N}(1+\hat{s_{j}}){\xi}_{kj}
+\frac{1}{4}\sum_{j=1}^{N}\hat{s_{j}}+\frac{N}{4}-b
\end{equation} 
From the constraints appearing in Eqs.(4) and (5), 
if we treat the case of $N/4{\ll}b$, we can put most of the items  
into the knapsack. 
On the other hand, the number of items 
to  be in the knapsack is too small when we set $N/4{\gg}b$.
For this  reason, we treat the case of $N/4=b$ which means that 
about half of the items go into the knapsack. 

The case of finite $K$ in the limit $N$  
infinity was investigated by Meanti \etal \cite{Mea} 
by the Lagrangian relaxation 
technique. 
They gave the explicit results 
for the special cases of $K=1$ 
and $K=2$.
Recently, Fontanari \cite{Fon} presented 
an explicit calculation 
of the annealed entropy of the configuration $
\mbox{\boldmath $s^{*}$}$ which satisfies $K$ constraints and 
gives the total benefit $E=\sum_{i}s_{i}^{*}/2$. He 
estimated the upper limit of the 
total benefit using the zero entropy condition.
  
He compared it with the exact result from 
Lagrangian relaxation for the case of $K=1,2$ (exact solutions) and $K>2$ 
to conclude that the annealed approximation becomes 
good as $K$ increases.
 
The original knapsack problem consists of 
only one constraint, or at most, several constraints.  
One may feel that only such a case is worth investigating.
However, in actual operations research field, we sometimes face
 multi-constraint ($K{\sim}N$) knapsack problems \cite{Vas}, an example of 
which is investigated in this paper.

If we look at the above problem from an actual operations 
research (OR) point of view,
difficulties lie in the discreteness 
 of knapsack variables. 
When hardship of this sorts confronts us, it is customary to  
apply the ``linear programing relaxation" technique \cite{Vas}. 
Linear programing relaxation is an approach to solve the integer 
programing problem approximately, 
which brings us to the 
neighborhood of the exact solution. 
Actually, for linear programing problems, a lot of useful 
methods, for example, the simplex method or the interior point method, 
have been proposed and improved \cite{Vas}. 

In addition, 
it is worth while to investigate the objective 
function for a linear programing problem itself  
(not as a relaxation of 
integer programing problem such as the orignal 
knapsack problem) \cite{Vas}. 
Linear programing problem is defined as follows
\begin{equation}
{\max}\left\{\frac{1}{2}\sum_{j=1}^{N}\hat{s_{j}}\Bigm|
\sum_{j=1}^{N}a_{kj}\hat{s_{j}}{\leq}b,
\mbox{}k=1,{\cdots},K,0{\leq}\hat{s_{j}}
{\leq}\infty,j=1,{\cdots},N \right\}
\end{equation}
Problems of 
the above style 
occurs very often 
in our life.
``The diet (nutrition) problem" is one of them. 
The diet problem ask us to determine the intake $\{s_{j}\}$ 
of each food. 
Each food $s_{j}$ has its own nutrients $a_{kj}$,
 {\rm i.e}, calcium,  protein, 
 vitamin, energy, etc.,
and one has to take  
each nutrient 
within the limit, 
{\rm i.e} 
$\sum_{j}a_{kj}{\leq}b$ 
$(k=1,{\cdots},K)$, 
where $a_{kj}$ is 
$k$th nutrient of 
the $j$th food. 
We should notice 
that valuables in 
this problem are real 
numbers, not integers.
And linear programing problem has more application 
than for integer 
programing problem \cite{Vas}.  
For two reason, it is worth investigating 
the multi-constraint ($K{\sim}N$) continuous 
knapsack problem.

When we look at our knapsack problem from the stand-point of 
linear programing relaxation, the knapsack valuable $s_{i}$, 
which has a real 
value and determines whether the $i$th item 
goes into knapsack, takes any value in the real subspace 
satisfies $\sum_{i}s_{i}^{2}=N$. 
We treat the next real-variable problem in the present paper.
\begin{equation}
{\max}\left\{U=\frac{N}{4}+\frac{1}{2}\sum_{j=1}^{N}\hat{s_{j}}\Bigm|
Y(\{{\bf s}\},\{a_{ki}\},b){\leq}0,
\mbox{}k=1,{\cdots},K,\hat{s_{j}}{\in}
\{-\infty,+\infty\},j=1,{\cdots},N \right\}
\end{equation}
and Eq.\,(5).
Although this relaxation of constraint $\sum_{i}\hat{s}_{i}^{2}=N$ and 
$\hat{s}_{i}{\in}\{-\infty, +\infty\}$ in Eq.(7)
somewhat 
obscures the direct significance of 
the variable $\mbox{\boldmath $s_{i}$}$, the global macroscopic behavior of 
the system is expected 
not to  be affected by this approximation 
as was mentioned above. 
And this constraint could be justified as simply the 
spherical version of the problem investigated by 
Korutcheva \etal \cite{Kou}

Using this linear programing relaxation technique, we can obtain 
 the approximate configuration 
$\mbox{\boldmath $s^{*}$}=(s_{1}^{*},{\cdots},s_{N}^{*})$ and we get 
higher value of the objective function (or the total profit) 
\begin{eqnarray}
E=\frac{1}{2}\sum_{j=1}^{N}s_{j}
\end{eqnarray} 
In this paper, 
we investigate 
the knapsack 
problem by the  
linear programing 
relaxation technique and 
estimate the objective function 
when the 
number of 
constraints 
is given for 
a large 
problem 
size.

\section{Replica symmetric theory}
For the parameter regions mentioned in the previous section, we consider 
the case with continuous variables under the  
normalization  constraint  $\displaystyle{\sum_{i=1}^{N}}(\hat{s_{i}})^{2}=N$.
For a large problem size,  
the fractional volume of solution $V(M,\{{\xi}_{ki}\})$ to Eq.\,(7) 
 is written by introducing the 
``fluctuation of the magnetization'' 
around $0$, which is of order ${\cal O}(1)$, 
\begin{eqnarray}
M=\frac{1}{\sqrt{N}}\sum_{i=1}^{N}\hat{s_{i}}
\end{eqnarray}
as
\begin{equation}
V(M,\{{\xi}_{ki}\})=\frac{\displaystyle{\prod_{k=1}^{K}}\int{d}\hat{s_{i}}{\Theta}\left(-(\displaystyle{\sum_{i=1}^{N}}
\frac{(1+\hat{s_{i}}){\xi}_{ki}}{\sqrt{N}}+\frac{M}{2})\right)\delta
\left(\displaystyle{\sum_{i=1}^{N}}(\hat{s_{i}})^{2}-N\right)\delta\left(\sqrt{N}M-\displaystyle{\sum_{i=1}^{N}}\hat{s_{i}}\right)}
{\displaystyle{\prod_{k=1}^{K}}\int{d}\hat{s_{i}}\delta\left(
\displaystyle{\sum_{i=1}^{N}}(\hat{s_{i}})^{2}-N \right)}
\end{equation}
This expression is similar to Gardner's volume of a binary 
spherical perceptron with local field 
$-\displaystyle{\sum_{i}}\hat{s}_{i}{\xi}_{ki}/\sqrt{N}$ and stability criterion constant $-M/2$ appearing 
in the problem of optimal capacity\cite{Gar} \cite{GD} or of 
the generalization of maximum stability rule \cite{Opp}. 
Thus it is possible to use the same technique.
And this $V$ is the value of configuration $\{\hat{s}_{i}\}$ which 
have a fixed $M$, implying also 
a fixed utility $U$. 
The typical value of $V$ is given by ${\exp}({\ll}{\log}V{\gg})$, 
where ${\ll}{\cdots}{\gg}$ denotes the averaging over quenched disorder. 
We perform the average of ${\log}V$ with the 
distribution $(3)$ over the different sets of solutions
 using the replica trick as follows.\\
\begin{equation}
{\ll}{\log}V{\gg}= \displaystyle{\lim_{n{\rightarrow}0}
\frac{{\ll}V^{n}{\gg}-1}{n}}
\end{equation}
The averaging of the power ${\ll}V^{n}{\gg}$ is accomplished 
by introducing an ensemble of $n$ identical replicas 
\\
{\footnotesize
\begin{equation}
{\ll}V^{n}{\gg}={\ll}\displaystyle{\prod_{\alpha=1}^{n}}
\frac{\displaystyle{\prod_{k=1}^{K}}\int{d}\hat{s_{i}}^{\alpha}{\Theta}
\left(-(\displaystyle{\sum_{i=1}^{N}}
\frac{(1+\hat{s_{i}}^{\alpha}){\xi}_{ki}}{\sqrt{N}}+\frac{M}{2})\right)\delta
\left(\displaystyle{\sum_{i=1}^{N}}(\hat{s_{i}}^{\alpha})^{2}-N\right
)\delta\left(\sqrt{N}M-
\displaystyle{\sum_{i=1}^{N}}\hat{s_{i}}^{\alpha}\right)}
{\displaystyle{\prod_{k=1}^{K}}\int{d}\hat{s_{i}}^{\alpha}\delta
\left(\displaystyle{\sum_{i=1}^{N}}(\hat{s_{i}}^{\alpha})^{2}-N\right)}
{\gg}
\end{equation}
}
Then we introduce the order parameter  which is the overlap between
two solutions  labeled by two replica indices ${\alpha}$ and ${\beta}$, 
\begin{equation}
q_{\alpha\beta}=\frac{1}{N}\sum_{i}\hat{s_{i}}^{\alpha}\hat{s_{i}}^{\beta}
\end{equation}
By the standard procedure, we obtain within 
the replica symmetric ansatz 
\begin{equation}
{\ll}V^{n}{\gg}=\exp\left[Nn\left\{\displaystyle{{\rm extr}_{q,\hat{E},\hat{q},\hat{M}}}G(q,\hat{E},
\hat{q},\hat{M})+{\cal O}(\frac{1}{N})\right\}\right] 
\end{equation}
where ${\rm extr}_{q,\hat{E},\hat{q},\hat{M}}$
 denotes the extremization with  respect to the parameters 
 $q,\hat{E},\hat{q}$ and $\hat{M}$.
Here 
\begin{equation}
G={\alpha}G_{1}+G_{2}-\frac{i}{2}\hat{q}q+i\hat{E}
\end{equation}
\begin{equation}
G_{1}={\log}{\hspace{.1in}}\displaystyle{\prod_{\alpha}}\int_{-{M}/{2}}^{\infty}\frac{d{\lambda}_{\alpha}}
{2\pi}\int_{-\infty}^{\infty}dx_{\alpha}\hspace{5cm}\nonumber \\
\mbox{}{\times}{\exp}\left[i\sum_{\alpha}{\lambda}_{\alpha}-{\sigma}^{2}\sum_{\alpha}
(x_{\alpha})^{2}-{\sigma}^{2}(1+q)\sum_{\beta}\sum_{\alpha<\beta}x_{\alpha}
x_{\beta}\right]
\end{equation}
\begin{equation}
G_{2}={\log}\int_{-\infty}^{\infty}\prod_{\alpha}
d\hat{s}^{\alpha}{\exp}\left[-i\hat{M}\sum_{\alpha}\hat{s}^{\alpha}
-i\hat{q}\sum_{\beta}\sum_{\alpha<\beta}\hat{s}^{\alpha}\hat{s}^{\beta}
-i\hat{E}\sum_{\alpha}(\hat{s}^{\alpha})^{2}\right]
\end{equation}
Using  the  saddle point equation with respect to $\hat{M}$, 
we obtain
\begin{equation}
\hat{M}=0
\end{equation}
Using this result and saddle point equation with respect to 
$\hat{E}$, we get
\begin{equation}
G(M,q)={\alpha}\int_{-\infty}^{\infty}Dt{\hspace{.1in}}{\log}
H\left(\frac{{M}/({2\sigma})-t\sqrt{1+q}}{\sqrt{1-q}}\right)+\frac{1}{2}{\log}(1-q)
+\frac{1}{2}\frac{q}{1-q}
\end{equation}
where
\begin{equation}
Dt{\equiv}\frac{{\exp}(-t^{2}/2)}{\sqrt{2\pi}}dt
\end{equation}
and
\begin{equation}
H(x){\equiv}\int_{x}^{\infty}Dt
\end{equation}
Finally we estimate the saddle point of Eq.(19) with respect to $q$ 
in the limit $q{\rightarrow}1$. This  means that only one optimal solution 
(which satisfies all $K={\alpha}N$ conditions in Eq.(7)) 
is selected for a given number of constraints in this limit 
and then optimal $M_{opt}$ can be calculated 
by   $q{\rightarrow}1$ for this optimal $\hat{s}$ 
which satisfies all $K$ constraints (7).
Thus we obtain $M_{opt}$ as a function of ${\alpha}=K/N$.
This result is shown  in figure $1$.

From this result, we see that $M-\alpha$ lines
 show the next behavior:  As the number of constraints increases,
$M$ decreases and as a result, total utilities  
 decrease (we should remember that total utilities are given as 
$U=N/4+\sum_{j=1}^{N}\hat{s}_{j}/2$ 
and $M=\sum_{i=1}^{N}\hat{s}_{i}/\sqrt{N}$). 
In figure 2, we also plotted the optimal profit $M_{opt}$ 
for the case of Ising variables using the zero entropy condition of 
solution discrete space taken from \cite{Kou}. 
This result shows that we can obtain the larger optimal profit 
than that of the original knapsack problem by 
our relaxation of variables.

\section{Three relevant lines}
\subsection{AT line}
In order to investigate the local stability of the RS solution, 
we follow the usual Almeida and Thouless \cite{Al} argument 
and introduce the fluctuations
 of the order parameters around the RS 
 order parameter as follows.
\begin{equation}
\begin{array}{ccc}
q_{\alpha\beta} & = & q+{\delta}q_{\alpha\beta} \\
{\hat{q}_{\alpha\beta}} & = & \hat{q}+{\delta}{\hat{q}_{\alpha\beta}}  \\
M_{\alpha} & = & M+{\delta}M_{\alpha} \nonumber \\
\hat{M}_{\alpha} & = & \hat{M}+{\delta}\hat{M}_{\alpha} \\
\hat{E}_{\alpha} & = & \hat{E}+{\delta}\hat{E}_{\alpha}
\end{array}
\end{equation}
It turns out that only fluctuations in $q_{\alpha\beta}$
and $\hat{q}_{\alpha\beta}$ lead to instability. 
The function to be investigated has the form
\begin{equation}
{\alpha}G_{1}(q_{\alpha}, M_{\alpha})+G_{2}(\hat{q}_{\alpha\beta},
\hat{M}_{\alpha}, \hat{E}_{\alpha})+i\sum_{\beta}\sum_{\alpha<\beta}
q_{\alpha\beta}\hat{q}_{\alpha\beta}
\end{equation}
From this expression, we can apply Gardner's \cite{Gar}\cite{GD} analysis to  our case and obtained AT-line as 

\begin{equation}
{\alpha}\left[\int_{-\infty}^{\infty}Dt
\left\{1-\frac{\int_{I_{z}}z^{2}Dz}{\int_{I_{z}}Dz}+
\left(\frac{\int_{I_{z}}zDz}{\int_{I_{z}}Dz}\right)^{2}
\right\}\right]^{2} < 1
\end{equation}
where 
\begin{equation}
\int_{I_{z}}{\equiv}\int_{({{M}/({2\sigma})-\sqrt{1+q}t})/{\sqrt{1-q}}}^{\infty}
\end{equation}
and  $q$ is the order parameter obtained by the replica symmetric calculation
and is given as
{\normalsize
\begin{eqnarray}
q={\alpha}\frac{\sqrt{1-q}}{\sqrt{1+q}}
\int{Dt}\left(\frac{M}{2\sigma}\sqrt{1+q}-2t\right)\hspace{1.6in}\nonumber \\
\mbox{}{\times}\frac{1}{\sqrt{2{\pi}}}\exp\left(-\frac{
({{M}/({2{\sigma}})-t\sqrt{1+q}})^{2}}{2(1-q)}\right){\Bigg /}H\left(
\frac{{M}/({2{\sigma}})-t\sqrt{1+q}}{\sqrt{1-q}}\right)
\end{eqnarray}
}
We plot this condition Eq.(24) in figure $1$.

From this figure, we see that the replica symmetry is stable for 
$M>M_{AT}=0.100$ ($\alpha<{\alpha}_{AT}=0.846$). 
However, when $M$ becomes smaller than  this critical value, 
the replica 
symmetry breaking occurs and the replica symmetric saddle 
point becomes unstable.
The AT point ($\alpha_{AT}(M_{AT})$) is a critical 
point where the replica symmetric order parameter $q_{RS}=q$ 
splits into (one step) replica symmetry breaking 
saddle point represented by $q_{0}$ and $q_{1}$. 
If this RSB transition at ${\alpha}_{AT}$ 
is continuous, the AT condition Eq.(24) gives a correct 
criterion of replica symmetry breaking.
On the other hand, if the RSB transition 
is discontinuous (first order), 
we should regard this discontinuity point as 
the symmetry breaking point 
rather than the AT stability limit. 
Fortunately, as we see in the next section, 
this transition is continuous and 
the AT argument is correct.

In order to understand this situation physically, we can investigate 
the replica symmetry breaking from the dis-connectivity
of the solution space $\hat{\mbox{\boldmath $s$}}$ according to 
Monasson and O'Kane \cite{Mon}, who treated RSB 
of perceptron with non-monotonic transfer function.\\
The solution condition Eq.(7) can be rewritten as 
\begin{equation}
\sum_{i}\hat{s}_{i}\frac{{\xi}_{ki}}{\sqrt{N}}{\leq}-\frac{M}{2}
+\sum_{i}\frac{{\xi}_{ki}}{
 \sqrt{N}}=-\frac{M}{2}+{\cal O}(1)
\end{equation}
Then if $M$ is positive, for many constraints $K$, 
 the space of $\hat{\mbox{\boldmath $s$}}$ which satisfies 
 Eq.(7) as well as the normalization constraint 
 $\sum(\hat{\mbox{\boldmath $s$}})^{2}=N$  consists of a 
 single domain and the solution space 
  is connected. On the other hand, 
  as $M$ becomes negative, the space of
$\hat{\mbox{\boldmath $s$}}$ splits into a lot of 
domains and the solution space 
 is disconnected. 
 
 Here we can estimate the ${\cal O}(1)$ term roughly 
 appearing on the right hand side of the 
 above inequality as $\sum_{j}{\xi}_{j}^{\mu}/\sqrt{N}\,{\sim}\,
 (1/\sqrt{N}){\times}{\sqrt{N}}{\sigma}=1/12$. 
 We should notice that this term is of ${\cal O}(1)$ and 
 we used the standard deviation ${\sigma}=1/12$.
 Therefore if we investigate the dis-connectivity of solution space, 
 replica symmetry must be broken for  
 $M<1/6=0.1667$. This value is not so far from the value obtained by 
 the AT 
 argument.
 We conclude that this dis-connectivity leads to 
 replica symmetry breaking.
 For this RSB region, we will later find the one 
 step replica symmetry breaking solution following the 
 scheme of Parisi \cite{Pari1}
 
\subsection{Zero entropy line}
In the previous subsection, we calculated the volume 
of solution space $\exp({\log}{\ll}V{\gg})$ for quenched random 
loads ${\xi}_{ki}$. 
For optimal $M_{opt}$, this volume 
shrinks to zero continuously.

Thus the increase in the number of constraints leads to a 
decrease of the solution space $\hat{\mbox{\boldmath $s$}}$  
in a discrete manner,  
and the entropy of solution 
space is $S(M_{opt})=0$ at $M=M_{opt}$. 
Following the idea suggested by Krauth and M$\acute{\rm e}$zard \cite{KM}
we expect that the optimal $M$ for continuous variables 
may be obtained  
when the volume of solution contains typically a single point
of hypercube. From analogy with the discrete-variable problem, 
this should occur around\\
\begin{equation}
V{\sim}\frac{1}{2^{N}}
\end{equation}
\subsubsection{Annealed calculation}
We first calculate the condition Eq.(28) by the annealed approximation as
\begin{equation}
{\log}{\ll}V{\gg}=G-H=-{\log}2
\end{equation}
This condition is easily calculated by 
\begin{equation}
G={\alpha}\,{\log}H\left(\frac{M}{2\sqrt{2}}\right)+\frac{1}{2}\left(1+\log(2\pi)\right)
\end{equation}
\begin{equation}
H=\frac{1}{2}\left(1+{\log}(2\pi)\right)
\end{equation}
Finally we get
\begin{equation}
{\alpha}\,{\log}H\left(\frac{M}{2\sqrt{2}}\right)=-{\log}2
\end{equation}
This is shown in figure $2$.\\
\subsubsection{Quenched calculation}
Next we calculate the above zero 
entropy condition Eq.(28) by the 
quenched calculation defined as 
\begin{equation}
{\ll}{\log}V{\gg}=\frac{1}{2^{N}}
\end{equation}
This leads to
\begin{equation}
{\alpha}{\int}{Dt}\hspace{.1in}{\log}H\left(\frac{{M}/{2{\sigma}}-t\sqrt{1+q}}{\sqrt{1-q}}\right)
+\frac{1}{2}\frac{q}{1-q}+\frac{1}{2}{\log}(1-q)=-{\log}2
\end{equation}
We plotted this result also in Figure 2.

From figure $2$ we find that this line Eq.(34) obtained by the 
quenched zero entropy condition is stable in the RS 
region.  
In figure 4 we plotted the behavior of the 
order parameter $q$ as a function 
of $\alpha$ which gives zero entropy and 
satisfies Eq.(34).
From these results, we may conclude that 
 our zero entropy approach gives a meaningful and consistent 
 criterion for the 
 optimal $M_{opt}$. 
\subsection{One step RSB solution}

In this subsection, we calculate the one step replica 
symmetry breaking solution following Parisi  \cite{Pari1}\cite{Pari2}. 
We try to find a first candidate
 for matrices \mbox{\boldmath $q$} and $\hat{\mbox{\boldmath $q$}}$ 
 appearing in the free energy.
 Using Parisi's suggestion, we divide  the $n$ replicas  into $m/n$ groups
  of $m$ replicas.
Next we set $q_{\alpha\beta}=q_{1},
\hat{q}_{\alpha\beta}=\hat{q}_{1}$, if $\alpha$ and $\beta$ 
belong to the same group, and $q_{\alpha\beta}=q_{0}, 
\hat{q}_{\alpha\beta}=\hat{q}_{0}$, 
if ${\alpha}$ and ${\beta}$ belong to 
different groups and we set $q_{\alpha\alpha}=0$.
We express \mbox{\boldmath $q$} and $\hat{\mbox{\boldmath $q$}}$ 
in terms of  a tensor product,
\begin{equation}
\mbox{\boldmath $q$}=(q_{1}-q_{0})\mbox{\boldmath $1$}_{{n}/{m}}
{\otimes}\mbox{\boldmath $e$}_{m}\mbox{\boldmath $e$}_{m}^{T}
+q_{0}\mbox{\boldmath $e$}_{n}\mbox{\boldmath $e$}_{n}^{T}
\end{equation}
\begin{equation}
\hat{\mbox{\boldmath $q$}}=(\hat{q}_{1}-\hat{q}_{0})
\mbox{\boldmath $1$}_{{n}/{m}}
{\otimes}\mbox{\boldmath $e$}_{m}\mbox{\boldmath $e$}_{m}^{T}
+\hat{q}_{0}\mbox{\boldmath $e$}_{n}\mbox{\boldmath $e$}_{n}^{T}
\end{equation}
where $\mbox{\boldmath $1$}_{k}$ is a $k$-dimensional unit matrix and $
\mbox{\boldmath $e$}_{k}^{T}=(1,{\cdots},1)$.
Using this form of broken symmetry of replica, 
we get the free energy as follows,
\begin{eqnarray}
f_{1RSB}(q_{1}, q_{0}, m)=\frac{\alpha}{m}
\int{Dy}{\log}\int{Dz}
\left\{
H\left(\frac{M/(2\sigma)-y\sqrt{1+q_{0}}-z\sqrt{q_{1}-q_{0}}}{
\sqrt{1-q_{1}}}\right)
\right\}^{m} \nonumber \\
\mbox{}+\frac{1}{2m}{\log}[1-q_{1}+m(q_{1}-q_{0})] 
\mbox{}+\frac{1}{2}(1-\frac{1}{m}){\log}(1-q_{1}) 
\mbox{}+\frac{q_{0}}{2[1-q_{1}+m(q_{1}-q_{0})]}
\end{eqnarray}
where we used the saddle point equations with respect to 
$\hat{q_{1}}$, $\hat{q_{0}}$, $\hat{E}$ and $\hat{M}$.

Before calculating the one step RSB solution, we had better 
investigate whether the one step RSB solution actually exists or not 
by minimizing the one step RSB free energy $f_{1RSB}$ 
with respect to the order parameters $q_{0}$ , $q_{1}$ and $m$.
For example, we try to investigate  
the case of $M_{opt}=0$. 
Varying ${\alpha}$ in the free energy $f_{1RSB}$, we look for 
 the set of order parameters $q_{0}$, $q_{1}$ and $m$ 
 to minimize $f_{1RSB}$. 
 In table 1, we list the result for the typical value of $\alpha$.
From this table, we see that until ${\alpha}\,{\sim}\,{\alpha}_{AT}=1.20$, 
the replica symmetric 
saddle point (which is $q_{0}=q_{1}=q$ and is independent of $m$) is
 stable and the one step RSB solution does not exist.
When we exceed the critical value ${\alpha}_{AT}$,
 the replica symmetric saddle point becomes unstable and 
 order parameter $q$ splits into one step RSB 
 order parameters $q_{0}$ and $q_{0}$ continuously as 
 shown also in figure 3. 
 From this continuous transition, we conclude that the 
 AT argument in the previous 
 section is consistent. 
 Comparing the free energy $f_{RS}$ and $f_{1RSB}$, 
 we see that the global minimum of this system 
 goes from the RS saddle point $q$ to the one step RSB saddle point 
 $(q_{0}, q_{1})$ at the AT instability point. 
 Therefore, the one step RSB solution actually exists.
 
In order to obtain  the solution in the limit 
$q_{1}{\rightarrow}1$, we use 
the scaling $q_{1}=1-{\epsilon}, q_{0}=Q$ and  $m={\mu}{\epsilon}$ to get 
\begin{equation}
F_{1RSB}(Q,\mu,\epsilon)=\frac{1}{\epsilon}
f_{1RSB}(Q,\mu)+O({\log}{\epsilon})
\end{equation}
Here
\begin{equation}
f_{1RSB}(Q,\mu)=\frac{\alpha}{\mu}
\int{Dy}
{\log}\left[H_{1}+H_{2}\right]+\frac{1}{2\mu}{\log}[1+\mu(1-Q)]+\frac{Q}{2[1+\mu(1-Q)]}\hspace{.7in}
\end{equation}
where
\begin{equation}
H_{1}{\equiv}H\left(\frac{{M}/({2\sigma})-y\sqrt{1+Q}}{\sqrt{1-Q}}\right)
\end{equation}
and 
\begin{equation}
H_{2}{\equiv}\frac{\exp(-\frac{\mu{(M/(2\sigma)-y\sqrt{1+Q})}^{2}}{2[1+\mu(1-Q)]})}{
\sqrt{1+\mu(1-Q)}}H\left(\frac{{M}/({2\sigma})-y\sqrt{1+Q}}
{\sqrt{1-Q}\sqrt{1+\mu(1-Q)}}\right)
\end{equation}
The saddle point equations  with respect to 
$\mu, Q$ and ${\epsilon}$ 
\begin{equation}
\frac{\partial F}{\partial \mu}\Bigg|_{\mu_{c}, Q_{c}, \alpha_{c}}
= \frac{\partial F}{\partial Q}\Bigg|_{\mu_{c}, Q_{c}, \alpha_{c}}
=F\Bigg|_{\mu_{c}, Q_{c}, \alpha_{c}}=0
\end{equation}
give the critical values ${\alpha}_{c}$ 
calculated by the one step RSB scheme.
We solved Eq.(42) numerically and show 
the result in figure 2.

From this figure, we see that the replica 
symmetry breaking decreases the optimal 
$M_{opt}$ slightly.
We also confirm this result from the 
fact that replica symmetry breaking point 
$q\,({\alpha}_{AT})$ comes 
close to $1$.
In figure 5, we also show the order parameter $q_{0}=Q$ as 
a function of the optimal profit $M_{opt}$. 
From this figure we see that, as $M_{opt}$ becomes close to $0.100$, 
{\rm i.e.} as replica symmetry breaking 
becomes weak, $Q$ becomes $1$ gradually
($Q=q_{0}=0.9912$ for $m=0.100$). 
This result is reasonable 
from the meaning of replica symmetry breaking. 
The reason is that in critical limit of ${\alpha}$,  
the one step RSB order 
parameter $q_{0}$ and $q_{1}$ becomes close to 
$q_{RS}=q$ and $q{\rightarrow}1$ 
if replica symmetry does not break down.

In order to confirm that this one step RSB solution is exact, 
we must check the stability of the one step 
RSB saddle point by the same technique as that used for the AT line 
 and investigate if  the free energy of one step calculation 
is lower than that of the second step calculation. Or the distribution 
of sizes of the ``disconnected'' domains of solution 
space \mbox{\boldmath $s$} must be computed analytically 
by  the same technique as in  Monasson and O'Kane \cite{Mon}.
This is a highly non-trivial problem. 
However we can conjecture that the two step replica symmetry 
breaking can hardly decrease the optimal profit even if two step RSB 
solution exists, 
because the optimal profit of replica 
symmetric calculation and that 
of one step replica symmetry breaking calculation are very close 
to each other as we saw above.
\section{Summary and Discussion}
We have calculated 
the optimal profit $M_{opt}$ explicitly for the case of 
continuous knapsack variables and the number of 
constraints $K$ is of the same order with the number of 
items $N$ by replica symmetric 
calculation explicitly.
From the AT argument, the replica symmetric solution 
becomes unstable for $M<0.100$.
We also investigated  this symmetry breaking point
 by minimizing the one step replica 
 symmetry breaking 
 free energy directly and 
 saw that this transition is continuous and 
 the AT line is valid. 
 For the argument of 
the problem of the optimal 
storage capacity of 
the perceptron 
 with non-monotonic transfer function investigated by Monasson and O'Kane \cite{Mon}, the physical meaning of this RSB can be understood as the dis-connectivity
of the solution space. 
Therefore, from the condition of 
dis-connectivity of solution space, we roughly estimated 
where replica symmetry breaking begins. 
For the RSB region, it was necessary to try to find a 
replica symmetry breaking solution according to Parisi's procedure \cite{Pari1}
 \cite{Pari2}. Thus using Parisi's scheme, we obtained the 
 one step RSB solution  
and the result shows that replica symmetry 
breaking makes the $M_{opt}$ decrease slightly.
Because it is not easy to confirm that 
this one step RSB calculation is exact, 
we can roughly expect that even if we calculate 
the two step RSB solution or three step RSB solution or  further 
step RSB solution, the optimal profit obtained at these further 
RSB saddle point should hardly increase. 
In stead of investigation in this direction, 
in the replica symmetric region, we can find a 
consistent solution 
using the zero entropy approach  and this result agrees with 
the discrete variables case \cite{Kou}
 for the most part.
This result may also support  the validity of their  
RS calculation.
The present  statistical mechanical approach is  expected to be applicable
the other linear programming  problems.  \\

\ack
I thank Prof. H. Nishimori for useful discussions  during this work.\\

\Figures
\Figure{ 
$M_{opt}-\alpha$ line calculated 
by the replica symmetric theory, 
AT line and one step replica symmetry 
breaking solution. 
In each line, we set ${\sigma}=1/12$. 
The replica symmetric solution is stable 
above $M\,{\sim}\, 0.100$.
In this figure,  the region where 
the replica symmetry is broken is shown enlarged. 
The optimal profit $M_{opt}$ decreases slightly by 
the one step replica symmetry breaking calculation. 
}
\Figure{
The zero entropy lines (quenched and 
annealed calculations).
The quenched zero entropy line is stable in the RS region. 
The optimal profit for Ising variables 
is also plotted.
This line is very close to the 
quenched zero entropy line. 
}
\Figure{
Order parameter $q$ as a function of 
the relative number of 
constraints ${\alpha}$ for the case of $M_{opt}=0$.
Replica symmetric order parameter 
$q$ splits into the one step replica symmetry breaking 
order parameter $(q_{0},q_{1})$ continuously 
at ${\alpha}_{AT}\,{\sim}\,1.20$.
}
\Figure{
Overlap $q$ which satisfies 
the zero entropy condition.
}
\Figure{
One step RSB order parameter 
$q_{0}=Q$ as a function of the 
optimal profit $M_{opt}$ in the limit of 
$q_{1}{\rightarrow}1$. 
As replica symmetry breaking becomes weak ( 
as $M_{opt}$ closes to $0.10$, ${\alpha}(M_{opt}=0.100)={\alpha}_{AT})$, 
$Q$ closes to $1$. 
}
\Tables
\begin{center}
\begin{tabular}{|c@{\quad\vrule width0.8pt\quad}c|c|c|c|c|c|}
\hline
$\alpha$ & $q_{RS}$ & $q_{0}$ & $q_{1}$ & $m$ & $f_{RS}$ & $f_{RSB}$ \\
\noalign{\hrule height 0.8pt}
$0.90$ & $0.7129$& ---------&---------&---------& $-1.5336$&---------  \\
\hline
$1.00$ & $0.7773$ &---------&---------&---------& $-1.9346$&--------- \\
\hline
$1.10$ & $0.8408$ &---------&---------&---------& $-2.4859$&--------- \\
\hline
$1.20$ & $0.8901$ &---------&---------&---------& $-3.3794$ &--------- \\
\hline
$1.25$ & $0.9073$ &  $0.8213$ & $0.9901$  & $0.1688$ &  $-3.7642$ & $-5.7847$ 
\\
\hline
$1.30$ & $0.9211$& $0.8155$ & $0.9902$  & $0.1858$ & $-4.4110$ & $-6.0763$\\
\hline
\end{tabular}
\end{center}
\Table{
The replica symmetric order parameter $q$ and 
the 1-step replica symmetry breaking order 
parameters  $(q_{0},q_{1},m)$ are listed  with the free energies 
for several values of 
$\alpha$.
Untill ${\alpha}{\sim}{\alpha}_{AT}=1.20$ 
the RS saddle point is stable and the 1-step 
RSB solution does not exist. 
Above the ${\alpha}_{AT}$, the free energy of 1-step RSB 
approximation is lower than that of RS approximation. 
}
\endtab


\Bibliography{14}
\bibitem{Kou}
Korutcheva E, Opper M  and  L$\acute{\rm o}$pez L  {\it J. Phys. A: Math. Gen.} {\bf 27} L645 (1994)
\bibitem{SK}
Kirkpatrick S and Sherrington D  {\it Phys. Rev.} B {\bf 17} 4384 (1978)
\bibitem{Mea}
Meanti M, Rinnooy Kan A H G, Stougie L and Vercellis C {\it Mathematical 
Programing} {\bf 46} 3 (1990)
\bibitem{Fon}
Fontanari J F  {\it J.Phys.A: Math.Gen.} {\bf 28} 4751 (1995)
\bibitem{Vas}
Chv$\acute{\rm a}$tal V {\it Linear Programing} (W.H.Freeman and Company 1983)
\bibitem{Gar}
Gardner E  {\it J. Phys. A: Math. Gen.} {\bf 21} 271 (1988)
\bibitem{GD}
Gardner E and Derrida B  {\it J.Phys.A: Math.Gen.} {\bf 21} 257 (1988)
\bibitem{Opp}
Opper M, Kinzel W, Kleinz J  and Nehl R  {\it J. Phys. A: Math. Gen.} {\bf 23}
 L581 (1990)
\bibitem{Zio}
Zionts S  {\it Linear and Integer Programming}   (John Wiley 1974)
\bibitem{Al}
de Almeida J R and Thouless D J   {\it J. Phys. A: Math. Gen.} {\bf 11} 983 (1978)
\bibitem{KM}
Krauth W and M\'ezard M  {\it J. Phys. France}  {\bf 50} 3057 (1989)
\bibitem{Mon}
Monasson R and O'Kane D  {\it Europhys. Lett.} {\bf 27} 85 (1994)
\bibitem{Pari1}
Parisi G  {\it J. Phys. A: Math. Gen.} {\bf 13} 1101 (1979)
\bibitem{Pari2}
M\'ezard M, Parisi G, Virasoro M  {\it Spin Glass Theory and Beyond}
 (World Scientific: Singapore  1986)
\endbib
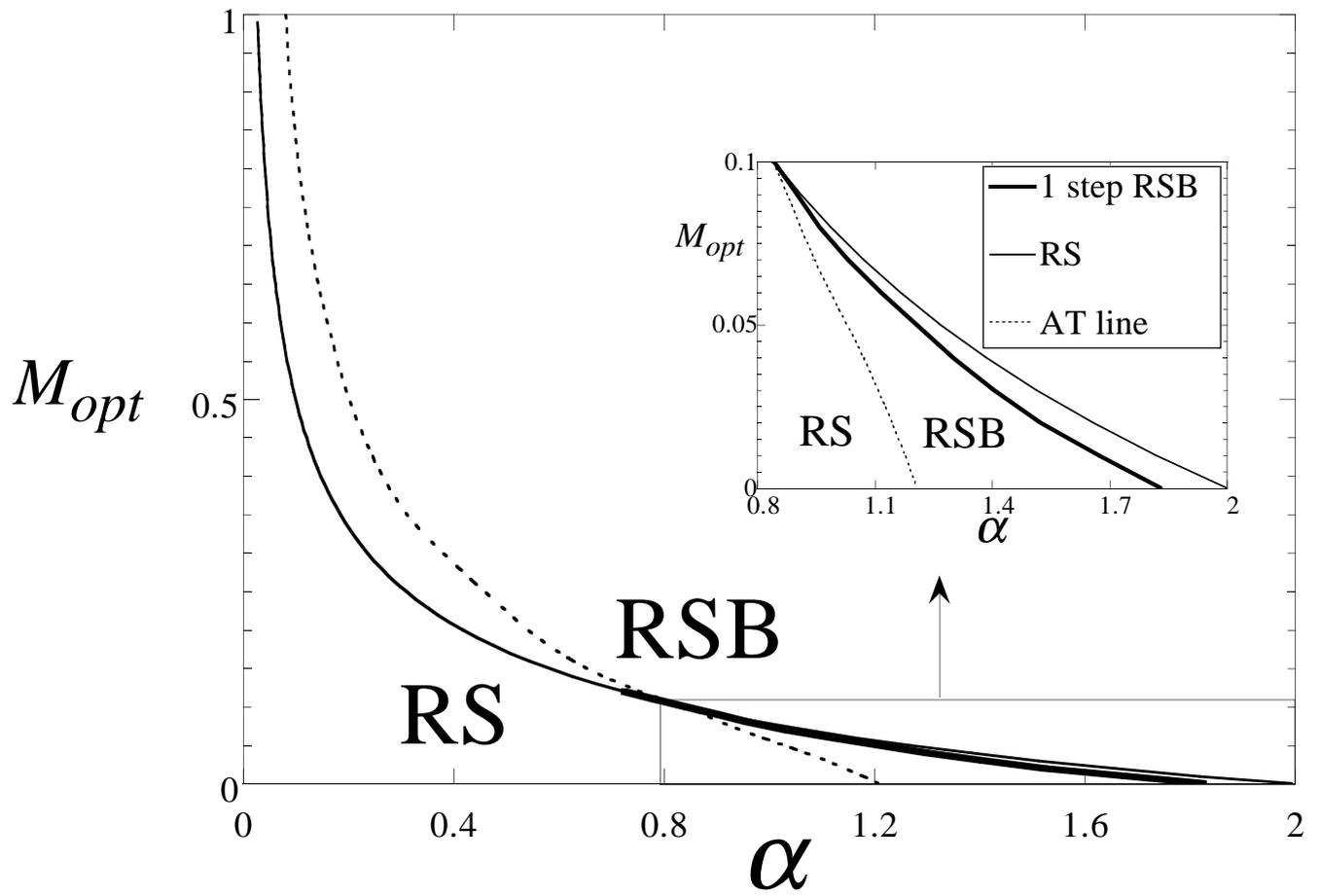
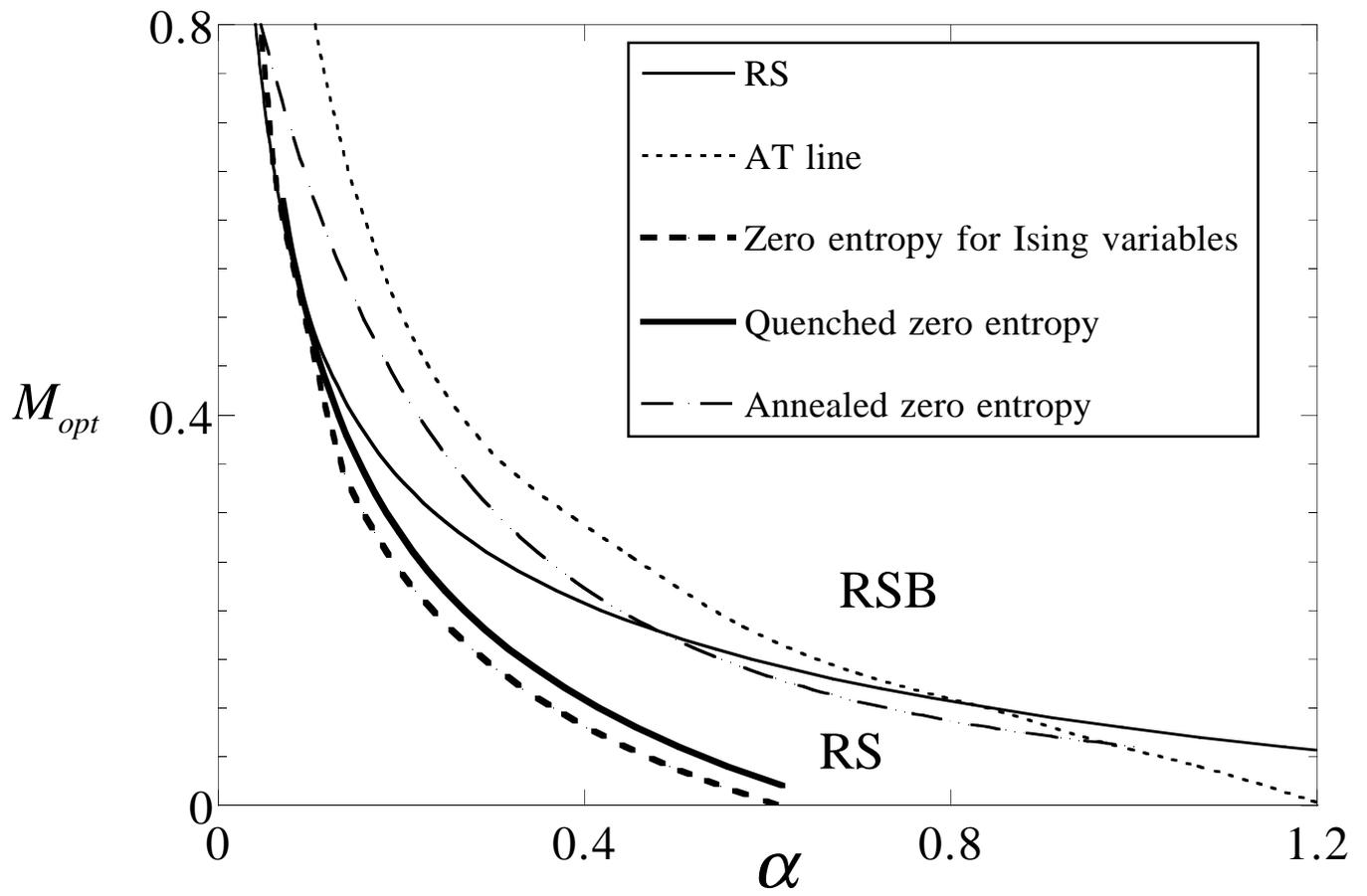
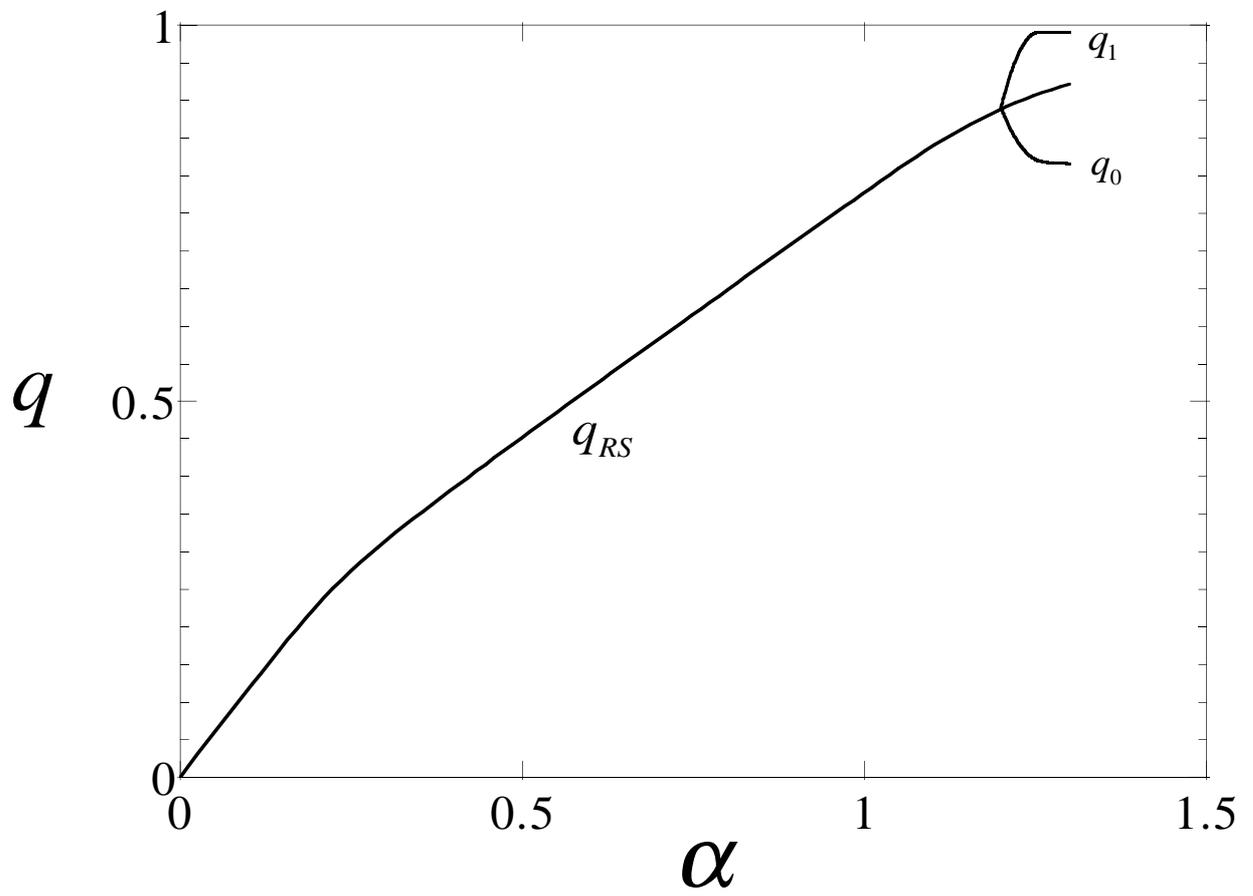
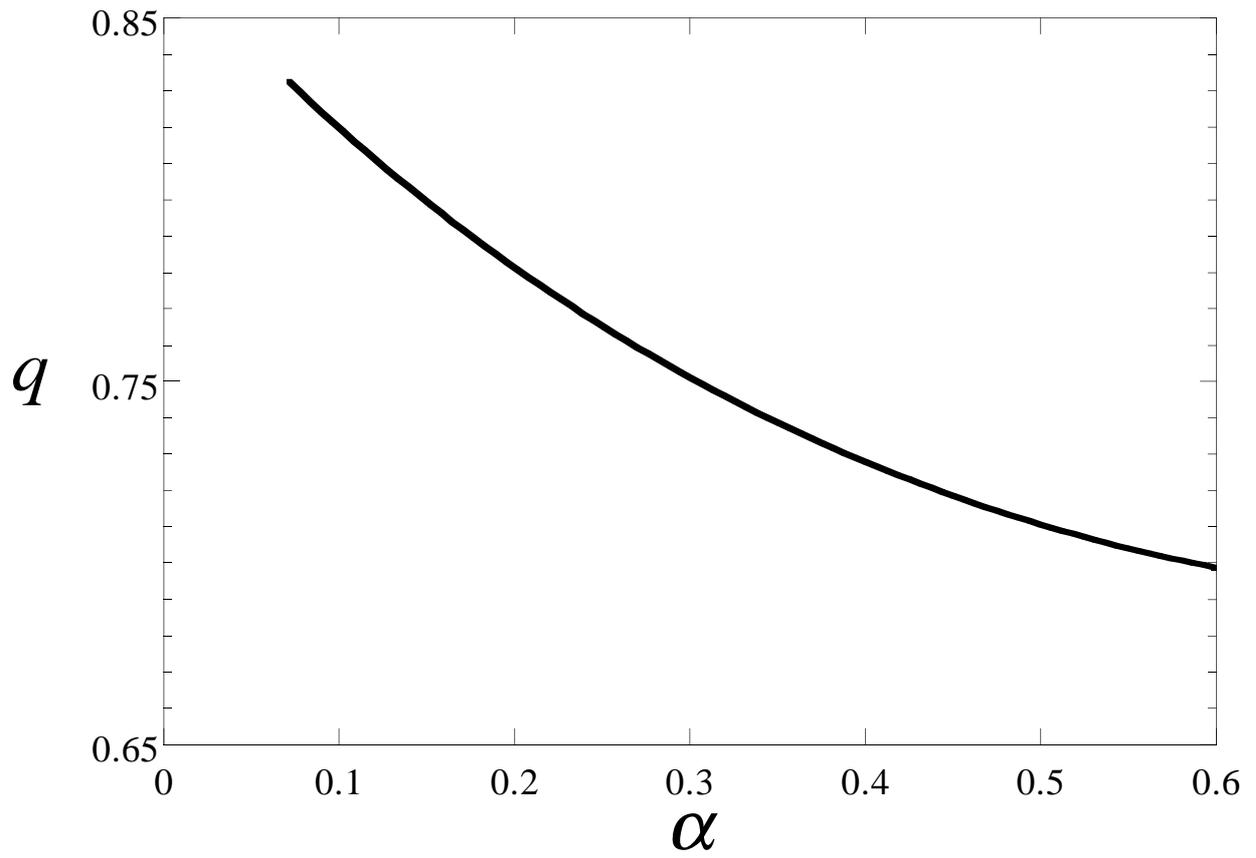
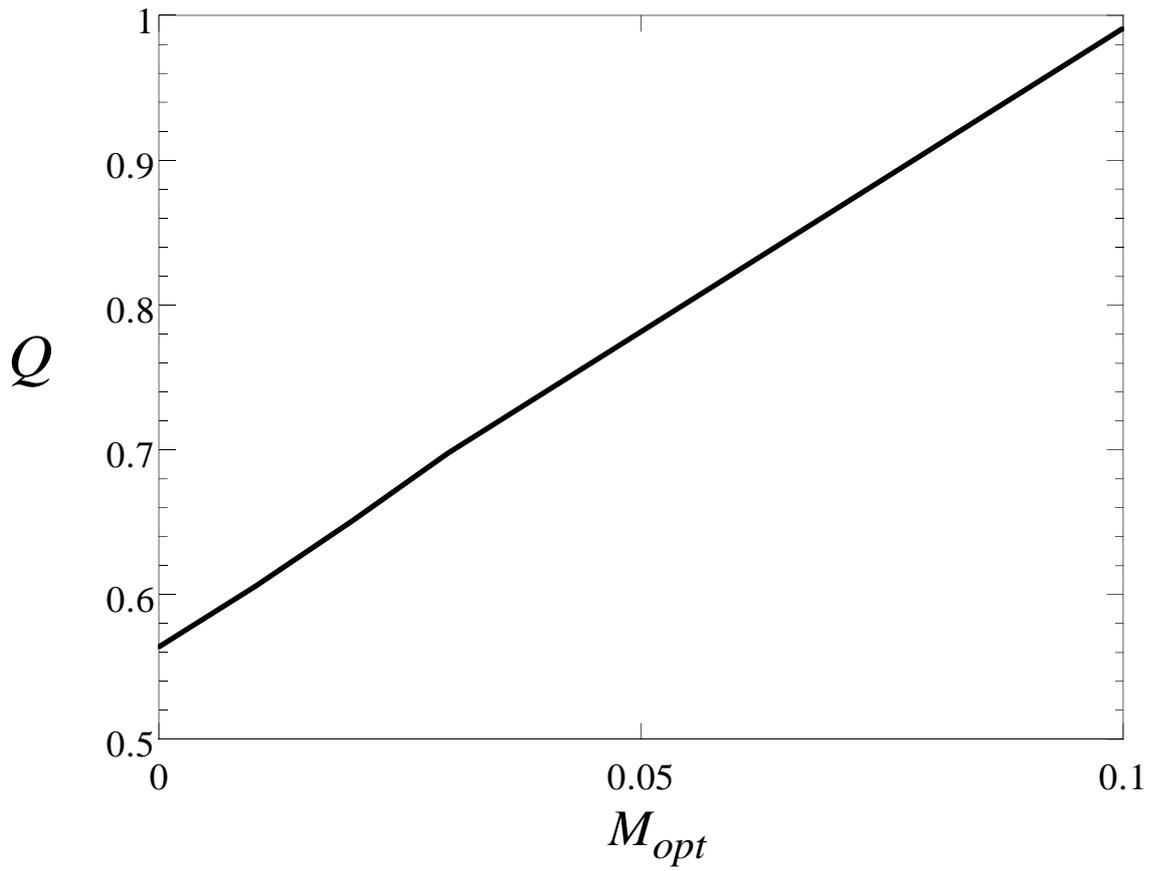
\end{document}